\PassOptionsToPackage{unicode}{hyperref}
\PassOptionsToPackage{hyphens}{url}
\documentclass[
]{article}
\usepackage[a4paper, total={6in, 8in}]{geometry}

\usepackage{wrapfig}
\usepackage{blindtext}
\usepackage{multicol}
\usepackage{lmodern}
\usepackage{authblk}
\usepackage{amsmath,amssymb}
\usepackage{url}
\usepackage{natbib}
\usepackage{xcolor}
\usepackage{amsmath,amssymb}
\usepackage{xurl}
\usepackage{iftex}
\ifPDFTeX
  \usepackage[T1]{fontenc}
  \usepackage[utf8]{inputenc}
  \usepackage{textcomp} 
\else 
  \usepackage{unicode-math} 
  \defaultfontfeatures{Scale=MatchLowercase}
  \defaultfontfeatures[\rmfamily]{Ligatures=TeX,Scale=1}
\fi
\usepackage{lmodern}
\ifPDFTeX\else
\fi
\IfFileExists{upquote.sty}{\usepackage{upquote}}{}
\IfFileExists{microtype.sty}{
  \usepackage[]{microtype}
  \UseMicrotypeSet[protrusion]{basicmath} 
}{}
\makeatletter
\@ifundefined{KOMAClassName}{
  \IfFileExists{parskip.sty}{%
    \usepackage{parskip}
  }{
    \setlength{\parindent}{0pt}
    \setlength{\parskip}{6pt plus 2pt minus 1pt}}
}{
  \KOMAoptions{parskip=half}}
\makeatother
\usepackage{longtable,booktabs,array}
\usepackage{calc} 
\usepackage{etoolbox}
\makeatletter
\patchcmd\longtable{\par}{\if@noskipsec\mbox{}\fi\par}{}{}
\makeatother
\IfFileExists{footnotehyper.sty}{\usepackage{footnotehyper}}{\usepackage{footnote}}
\makesavenoteenv{longtable}
\usepackage{bookmark}
\IfFileExists{xurl.sty}{\usepackage{xurl}}{} 
\hypersetup{
  hidelinks,
  pdfcreator={LaTeX via pandoc}}

\providecommand{\keywords}[1]
{
  \small	
  \textbf{\textit{Keywords---}} #1
}

\title{Other Worlds: Using AI to Revisit Cybersyn and Rethink Economic Futures}

\author[1]{Luke Munn}
\author[2]{Liam Magee}
\affil[1]{University of Queensland, Australia \authorcr l.munn@uq.edu.au}
\affil[2]{University of Illinois Urbana Champaign, United States \authorcr lmagee@illinois.edu}

\date{November 2024}

\begin{document}

\maketitle

\begin{abstract}
Neoliberalism has become orthodoxy in the present, erasing competing
paradigms and alternative imaginings. Chile's radical Cybersyn project
from 1971 to 1973 offers a departure point for an alternative path,
albeit one that was abruptly and violently extinguished. We revisit this
moment by fine-tuning AI language models on the words and writing of
Salvador Allende, the Chilean President, and Stafford Beer, the
cyberneticist who helped to design the project. We conduct interviews
with these simulated personas, focusing on how their revolutionary ideas
might be taken up in the present. We then use an AI model to generate
five-year-plans from 1973 to the present, simulating an alternate
history guided by Cybersyn and a progressive agenda. We frame these
interventions as \emph{socialist infrastructuring} that cultivates a
more expansive \emph{socialist imagining}. This work is not about the
viability of planned economies, but about the ``inspirability'' of
exploring other value-systems in the present, allowing us to break out
of our future-on-rails to envision alternative ways of organizing
economy and society.
\end{abstract}

\keywords{computation, Cybersyn, socialist calculation, neoliberalism, machine
learning, AI, futures}

\pagebreak

\begin{multicols*}{2}

\subsection{\texorpdfstring{\textbf{Introduction}}{Introduction}}\label{introduction}

In the last four decades, neoliberal policies have ascended to dominate
economic, environmental, and education decision-making, marginalizing
alternative conceptions \citep{hursh2011}. Hegemony is achieved
not through the result of any coordinated campaign, but rather through a
far more ambient constellation of claims, norms, and activities diffused
throughout social and political institutions and everyday life. When
critique does arise, it is quickly defused through co-option \citep{boltanski2017}.

The success of this colonization of imagination has meant that these
free market ideologies attain the status of common sense. The neoliberal
status quo becomes stabilized as something which is normal or natural
\citep{davis2022}. The result of this overwhelming dominance is capitalist
realism \citep{fisher2010}: the sense that capitalism is the only viable
political and economic system and it is impossible even to imagine a
coherent alternative to it. As \citep{ghosh2017} stresses, our contemporary
crises are only exacerbated by our crisis of imagination. The need to
envision real utopias, sufficiently articulated alternatives to
unbridled capitalism, is more urgent than ever \citep{wright2010}.

How, then, to break out of this incontestable frame, to fracture this
commanding but extremely narrow mode of understanding the world? One
approach is to draw upon other times and places. The past is a foreign
country \citep{lowenthal2015}, a terrain whose inhabitants think and act
differently. Indeed, the pervasive commodification and homogenization of
contemporary life means the past becomes alien \citep{strasser2003}, a way of
life that seems unfamiliar or even absurd. Yet for precisely this
reason, the past contains enormous potential, a set of alternative
values and visions that can be leveraged to expand our conceptual
horizons in the present.

Our point of departure is Cybersyn, a project undertaken in Chile
between 1971 and 1973. Inspired by both cybernetics and socialist
principles, Cybersyn aimed to use computation to model, organize, and
optimize a nationalized economy. Though flawed in particular ways \citep{medina2011}, the project was also ambitious, innovative, and
politically progressive, forming a moment of radical potential \citep{morozov2023a}. Our article thus joins other scholarship, special issues (Alvarez
and Gutierrez 2022), and exhibitions (Palmarola, Medina, and Alonso
2024) that have recently revisited this moment.

More broadly, our exploration aligns with investigations into analogous
moments of alterity, such as AI development in the late socialist GDR
\citep{schmitt2023} or economic cybernetics in Soviet Russia \citep{gerovitch2004,
peters2017}. While these are historical investigations first and
foremost, they more implicitly register a dissatisfaction with the
de-facto convergence between capitalism and computation in the West, and
point to other approaches and possibilities. Following this thread,
Phillips and Rozworkski's \citep{phillips2019} sardonically titled \emph{The People's
Republic of Walmart} is one recent popular text that links both Soviet
and Cybersyn historical cases to the algorithmic economic planning
exercised by today's multinational corporations -- and to alternative
configurations of ownership and control such links imply.

To revisit Cybersyn, we carry out two interventions. First, we simulate
its two key architects, Salvador Allende and Stafford Beer, by
fine-tuning AI language models based on their words and speech. Second,
we use AI language model agents to generate an alternative history where
Cybersyn continued and informed policy decisions up until the present.
Grappling with these technologies can be generative conceptually and
theoretically---a form of \emph{socialist infrastructuring} that fosters
\emph{socialist imaginings}. The lives of Allende and Beer and the
alternate history of a functioning Cybersyn are social sites, with paths
both actualized and untaken, that point towards alternative futures or
that which ``might have been'' \citep{palmer2014}.

Like other ``economic science fictions'' \citep{davies2019}, these concepts and stories harness the power of the utopian imagination
to revitalize economic thinking. These interventions are thus not about
proving the \emph{viability} of socialist calculation, but about
leveraging it for its \emph{inspirability}, generating new questions and
approaches to longstanding sociocultural, political, and financial
problematics. They help us to see, in small but significant ways, how
these radical ideas might inform our present conditions and allow us to
break out of our future-on-rails. In the sections below, we first
provide basic context on Cybersyn, then step through our two
interventions with AI, discuss them as socialist infrastructuring and
imagining, and conclude by articulating their contribution.

\subsection{\texorpdfstring{\textbf{Context: Cybersyn Then and
Now}}{Context: Cybersyn Then and Now}}\label{context:-cybersyn-then-and-now}

CyberSyn was a cybernetics-based project implemented in Chile between
1971 and 1973 under the presidency of Salvador Allende. Allende's
campaign had been radical, promising to free Chile from capitalism and
imperialism. Upon coming to power in 1970, Allende immediately launched
a sweeping programme of reforms based on his longstanding commitment to
socialism aimed at addressing social inequities, including the
nationalization of industries, worker participation, and land reforms,
amongst others.

CyberSyn aimed to optimize the coordination and efficiency of national
industries by integrating data from various sectors and providing
decision-makers with timely information. To carry out this task,
Allende's government hired the British cybernetician, Stafford Beer.
\citep{beer1984, beer1993, beer1995} championed the use of his viable systems model
for both public and private sector clients, developing models which
aimed to map the key components of a firm and optimize these processes.
Beer traveled several times to Chile, working closely with key
government stakeholders to design and implement a cybernetic model.

The Cybersyn project essentially consisted of four elements:
\emph{Cybernet}, a network of telex machines distributed across the
country in state-run enterprise offices that would feed information back
to headquarters; \emph{Cyberstride}, statistical modeling software that
would visualize and model this performance and attendance data;
\emph{CHECO}, simulator software intended to model possible results of
economic decisions; and \emph{Opsroom}, a special room designed for
administrators that would allow them to view this information on
screens, make informed decisions, and communicate these decisions back
to enterprises and factories \citep{medina2011}.

Computation was key to this vision. Although the ``technocratic'' is
today associated with neoliberalism, Cybersyn's designers believed
technology could ``extend collective politics in an entirely
anti-neoliberal sense.'' The system aimed to collect a wide
range of data regarding production, distribution, and consumption,
integrate it into an all-encompassing and dynamic model, and then use
this model to simulate the economy and anticipate future issues,
supporting adaptive decision-making. To support this goal, the
government attained the first mainframe computer in Latin America and
established a national network of 500 telex machines. However, for both
\citep{beer1993} and Allende \citep{debray1971}, such computation was not merely
technical but social in seeking to construct a more equitable economy
that fostered worker autonomy and workplace democracy.

As \citep{medina2011} chronicles, these aspirations were only partially
implemented, with the project running up against technical, financial,
and social frictions, as well as a broader backlash due to rising
inflation and US-supported worker strikes. However the project only ran
for a brief period. Within two short years, Allende's presidency was
violently cut short by a CIA-backed military coup, leading to his death
and the installment of the Pinochet regime.

Despite its relatively short existence, CyberSyn left a lasting legacy
as an innovative experiment in leveraging computation and cybernetic
principles to socialist economic planning. Medina's \emph{Cybernetic
Revolutionaries} \citep{medina2011} chronicled the project in unprecedented depth,
achieving accolades and renewing interest in this unusual moment in the
history of technology. More recently, there has been a surge of
scholarly work reflecting on the origins, lessons, and continued
relevance of Cybersyn \citep{gill2022, vehlken2022, alvarez2022,
espejo2022}. In 2023, influential technology critic Evgeny
Morozov \citep{morozov2023a} also produced a podcast series exploring the Cybersyn
project. The podcast's byline, ``the tech world that may have been,''
gestures to the rich technical, social, and political horizons that the
project still opens up. Without overly idealizing the project or getting
mired in historical detail, we take inspiration from this recent
scholarly and popular interest. We see the Cybersyn moment as a key
inflection point for an alternative history and one that can be
revisited to expand our imaginative capacities in the present.

\subsection{\texorpdfstring{\textbf{Intervention 1: Resurrecting
Cybersyn's Architects}
\{\#intervention-1:-resurrecting-cybersyn's-architects\}}{Intervention 1: Resurrecting Cybersyn's Architects \{\#intervention-1:-resurrecting-cybersyn's-architects\}}}\label{intervention-1-resurrecting-cybersyns-architects-intervention-1-resurrecting-cybersyns-architects}

In the first of our interventions, we customized several large language
models (LLMs) based on texts from both Allende and Beer. In generating
these synthetic personas, we not only aimed to replicate these radical
figures from the past, but enable them to be dialogued with in order to
consider alternative approaches to contemporary issues in technology,
economy, and society.

To produce these models, we used a technique called fine-tuning, in
which a new dataset is used to adjust the parameters of a pre-existing
model. A generic ``foundation'' model such as GPT, for instance, might
be fine-tuned on hospital documents to create a model that can better
understand and respond to medical cases. We opted to fine-tune Llama 2
7b \citep{touvron2023}, an open source LLM released by Meta in June 2023. We
deliberately chose this model due to its size, flexibility and
capability to respond meaningfully to varied questions and tasks.

To construct the Allende model, we used \emph{Conversations with
Allende} \citep{debray1971}, a two-hundred-page interview with the Chilean
president. The interview format, with its question and answer structure,
maps neatly onto the prompt and response data structure that is often
used in fine-tuning models. To construct the Beer model, we used a
variety of his prolific written material on cybernetic systems,
including speeches, academic articles, and monographs, a dataset
totalling over three thousand paragraphs \citep{beer1983, beer1993, beer1995}. This data was structured using a blank prompt and paragraph
response, an alternative pattern recommended for some fine-tuning
applications \citep{mayne2023}. Fine tuning took several hours on a high
performance computer run in an author's office.

We carried out ``interviews'' or conversations with these models using a
standard chat interface. On a technical level, we were interested in
whether our infrastructuring work (detailed in the previous section) had
sufficiently honed a model so that it responded in ways similar to
Allende or Beer. On a conceptual level, we wanted to draw out their
radically different approach to society and economy and ask how this
vision might be applied today. Below, we present snippets from these
conversations and a brief analysis of each.

\emph{tell me about your work in Chile}\\
\emph{``Allende'' model: I served as a member of the Central Committee.
I was one of those who attended the Congress that founded the Party, and
I took part in all its activities.}

\emph{tell me about your work in Chile}\\
\emph{``Beer'' model: The first thing to say is that the whole of this
story was told by Allende himself, and published as a book. It has been
widely read; it is not my intention to give an account of his life or
death which would be other than what he wrote. But I do want to draw
attention to some aspects of the story which have been misunderstood.}

The first key finding is that we see a shift from generic to specific.
Asking a generic large language model about work in Chile would
inevitably trigger a set of clarifying questions: what work, by whom, at
what time? Instead, we see concise answers written from the perspectives
of Allende and Beer. This confirms that the fine-tuning was successful
in transforming the model from something universal to something which
appears to be much more individual, grounded in a specific subject,
time, and place.

We then asked a series of questions focused on applying their ideas to
the present. For the sake of space, we present a small selection of
these below.

\emph{what is the role of socialism today?}\\
\emph{``Allende'' model: The present world order and its defence are
condemned for their selfishness, exploitation, violence, oppression and
discrimination. Socialism offers mankind another way forward.}

\emph{how would you use computers today to accomplish this goal?}\\
\emph{``Beer'' model: I think that the most important thing about the
computer revolution for us was its potentiality to create a new kind of
organization - one based on information rather than authority.}

The concept of ``information rather than authority,'' to take just one
example, is compelling in seeing technical transformation as a way to
unlock new socioeconomic orders. The state, in the Weberian model, is
characterized by a monopoly on violence, an entity who holds the sole
right to the legitimate use of force \citep{weber1918}. ``Information rather
than authority'' gestures, albeit obliquely, to a different kind of
governance, a bottom-up rather than top-down model where shared
information provides a kind of common foundation for decision-making.
These visions of computation in the service of human flourishing echo
earlier socialist visions \citep{guevara1963} which aimed to ``turn machines
into liberating instruments'' that could achieve the most important
outcome: ``individuals developed to the full.''

While the ideas of these radical individuals can certainly be accessed
through their writings, the development of a bot or simulated persona
offered a different kind of engagement, a liveness \citep{auslander2002} that
enables conversation to unfold and critical ideas to be considered more
directly in the context of contemporary issues. The impact of this
intervention, then, is not about the responses in themselves, but about
the ability to ask a specific question of a simulated historical figure
like Allende or Beer using natural language and receive a conversational
response. These responses, in turn, offer a kind of springboard or
catalyst for further reflection, imagination, and exploration.

\subsection{\texorpdfstring{\textbf{Intervention 2: Generating an
Alternate Cybersyn-Directed
History}}{Intervention 2: Generating an Alternate Cybersyn-Directed History}}\label{intervention-2:-generating-an-alternate-cybersyn-directed-history}

In our second intervention, we use an AI-model to generate an alternate
history. Employing the ``what if?'' question of alternate history and
science fiction \citep{rosenfeld2002, wright2010, davies2019} we simulated
what may have occurred if Cybersyn continued, with sufficient data and
investment, to shape national decision-making and policy formation. We
used the smallest variant of Llama v3.1, with eight billion parameters,
to recreate the ``lofi'' sensibility of the original Cybersyn project:
within this fictional account, an entire national economy is ``planned''
on a Macbook laptop, and the unpredictable responses of this smaller
model also echo the errors of Cybersyn.

For the simulation, we create two LLM-based agents. The first acts as a
generator of history events from 1973 until the present. For each year,
this agent generates a summary of the global macroeconomic,
geopolitical, and technological situation. To provide further context,
we also feed global World Bank data, such as population, life
expectancy, GDP, and Gini coefficients, into the simulation. This helps
to correct potential distortions from presentations of crises.

The second, which we name ``CyberSim,'' roleplays as Cybersyn,
formulating policy responses to those macroeconomic events. Importantly,
CyberSim is prompted to make refinements to itself, commensurate with
changes in the technological environment: the rise of the
microprocessor, the personal computer, the Internet, Big Data, AI, and
so on. Cybersim's planning recommendations are ``presented'' to the
first agent, which operates from an ``orthodox'' economic perspective
and describes what it sees as the likely outcomes of these decisions.

\begin{table*}[ht]
  \centering
  \caption{Key phrases from CyberSim recommendations across selected periods}
  \begin{tabular}{p{0.5\linewidth} p{0.5\linewidth}}
  \toprule
  \textbf{Year Range} & \textbf{Key Phrases} \\
  \midrule
  1973-1978 & "export-oriented technologies", "comprehensive training and education" \\
  1978-1983 & "increase minimum wage", "increase healthcare spending" \\
  1983-1988 & "participatory decision-making", "innovation and entrepreneurship" \\
  1993-1998 & "government databases", "national cybersecurity", "e-learning platform" \\
  1998-2003 & "community-led initiatives", "participatory governance", "startup companies and innovation hubs" \\
  2003-2008 & "eco-tourism and sustainable development", "GDP growth rate", "public-private partnerships" \\
  2018-2023 & "enhance forecasting", "renewable energy", "national e-commerce platform" \\
  \bottomrule
  \end{tabular}
  \end{table*}

The table above presents key phrases from this simulation, which we
unpack in more detail here. The first two five-year periods (1973-78,
1978-83) are dominated by the aftermath of the 1973 oil crisis. In the
first period, CyberSim responds with policies that aim to diversify
energy sources, boost ``export-oriented technologies'' and undertake a
national ``comprehensive education and training program.'' In the second
period (1978-83), CyberSim amends these policies to focus on price
controls, decentralized decision-making, increased minimum wage, and
healthcare. Cybersyn will also be upgraded with ``new personal
computers,'' ``a network of computer terminals,'' and ``regular training
and workshops'' designed to increase knowledge and promote
``decentralized decision-making.''

The following periods (1983-88, 1988-93, 1993-98) adopt very similar
prescriptions. By the 1990s, the simulation suggests ambitious plans to
integrate ``government databases'' with Cybersyn, develop ``electronic
government'' initiatives, and address cybersecurity threats. None of
these initiatives are necessarily surprising -- they are common to
actual policy directives from the same periods of time -- but they are
continuously framed within the Cybersyn imperatives to increase public
engagement and involvement in decision-making. However, over successive
periods the simulation also becomes more focussed on economic and
technocratic planning, and the earlier emphasis on community becomes
more gestural. For example, the period 1998-2003 includes a goal to
``Strengthen Democratic Governance and Participation,'' with steps that
include ``community-led initiatives to promote participatory
governance.'' But in the following period (2003-2008) the focus returns
to economic and environmental objectives: policies include promoting
eco-tourism and sustainable development, and strengthening
public-private partnerships. By the final period (2018-23), Cybersyn
will become integrated with AI to ``enhance forecasting and predictive
modeling capabilities''; energy production will become decentralized,
incorporating solar and wind; and plans for ``participatory budgeting''
sit alongside designs for a ``national e-commerce platform.'' Schemes to
coordinate national economic activity through decentralized units of
production -- an integral part of the Cybersyn vision -- have
disappeared entirely.

We ran the simulation multiple times, with similar effects. In each
case, we see the model's willingness to blend the jargon of a given
decade with the overall socialist futurist agenda that we had originally
assigned. Rhetoric from the Cybersyn techno-socialist project is spliced
together with terms taken from more recent discourses of smart cities,
sustainable development, and participatory budgeting \citep{kitchin2015,
halpern2017, peck2015, hope2021}. This tendency is
clarified by thinking about policy recommendations as a dual move:
privileging some values and visions inherently means ignoring or
discarding others. The ambitious plans of Allende \citep{debray1971} -- a
flatter society, worker power, national autonomy -- are quickly left
behind, and the focus turns to economic growth, business innovation, and
capacity building. In other words, Cybersyn's radical socialist agenda
becomes diluted into a more general discourse of recent governmental
policy that seeks to merge progressive ideals with market-friendly
policies (``sustainable development'', ``public-private partnerships'',
and so on).

Certainly the simulation's suggestions are not inherently incompatible
with a technocratic socialist state. However, this language and framing
seems drawn from a neoliberal playbook rather than the more radical or
alternative strain of politics embodied in Cybersyn. Here the language
model reflects \citep{jameson1994} oft-quoted (and perhaps misattributed)
phrase that it has become easier to imagine the end of the world than
the end of capitalism. The early radicalism of the 1970s becomes
homogenized into a kind of bland progressivism even when the model is
given free reign to invent alternatives in this counterfactual account.

Our hypothesis, informed by the deep links between data and model in AI
systems \citep{ntoutsi2020, artelt2024}, is that these models are influenced by orthodox rather than heterodox economic
literature. In other words, training data is dominated by capitalist or
neoliberalist ideas, phrases, and case studies. Socialism effectively
becomes a ``low resource'' socioeconomic system, just as many
non-English dialects with a lack of material online are seen as ``low
resource'' languages by AI developers \citep{tonja2022,
hasan2024}. As the simulation marches towards the present, socialism is
diluted into centrist language that marries poverty reduction,
sustainable development and participatory budgeting with market-friendly
trade policies and technocratic control. In effect, the ``triumph'' of
the neoliberalist paradigm historically feeds back into the model's
training data, diluting alternative systems over time or erasing them
altogether.

\subsection{\texorpdfstring{\textbf{Discussion: Socialist
Infrastructuring for Socialist
Imagining}}{Discussion: Socialist Infrastructuring for Socialist Imagining}}\label{discussion:-socialist-infrastructuring-for-socialist-imagining}

Simulating ``Allende'' and ``Beer'' and generating an alternative
socialist history are technical but also political endeavors, striving
to broaden our imaginative horizons beyond the current socioeconomic
order. So if this material engagement is heterodox, it seems to be a
fitting way of considering heterodox economics.

Fine-tuning an AI model is just one possible manifestation of a broad
array of practices that we term \emph{socialist infrastructuring}.
Socialist infrastructuring aims to contest capitalist dominance by
developing tools, systems, and software that operate according to an
alternative set of values predicated on shared ownership and the
communal management of economic, social, and ecological resources. Such
infrastructures strive to cultivate collective equality and solidarity,
values typically ignored or even stigmatized in contemporary life.

While these methodological descriptions appear cut-and-dried, in reality
this infrastructuring work was far more messy. The rate of change in
this AI space is extremely fast, with new models and architectures
released on a weekly or even daily basis. A vast amount of software
libraries, tools, and code frameworks have been rapidly developed to
work with these models. However, these tools are both immature, taking
the form of alpha or beta releases, and quickly obsolescent. The result
is often broken code, deprecated functions, and libraries plagued by
incompatibilities. For both authors, this meant days of ``configuration
hell,'' trial-and-error, and filing bug reports, before tools could be
properly used to carry out our aims.

Such infrastructuring work continues a tradition of exploration through
the design and development of alternative systems. Cybersyn leveraged
everyday (and indeed outdated) technology -- a mainframe, a network of
telex machines, a set of projectors -- to develop a compelling portrait
of a different economy \citep{medina2015}. And these systems might not
necessarily be composed of chips, cables, and electricity. In later
years, Beer hypothesized that the management of a highly complex system
like the economy might best be carried out by a similarly complex
ecological system: in response he established a pond, populated it with
organisms and microorganisms, and documented the results. As \citep{pickering2010} notes, using a pond as a basis of factory management is a radical
idea driven by a very different kind of perspective to our modern
technoscientific one. Implicit in these experiments is a sense that
orthodox approaches are insufficient and that ecosystems might ``solve''
for complexity in a more efficient or appropriate (albeit invisible)
way.

Socialist infrastructuring paves the wave for \emph{socialist
imagining}. By socialist imagining, we mean the ability to envision,
conceptualize, and communicate alternative ways of organizing the
socio-economic order which depart from neoliberal and late-capitalist
logics and follow more communal and sustainable principles. If the
market, contrary to established consensus, is not the only mechanism
capable of resolving the immense complexity of the economy, what other
paradigms, logics, or approaches might be used to efficiently---and
equitably---allocate resources? Or even more broadly, picking up on
rhetoric from Beer and Allende, what is the purpose of an economy? How
might an economy be structured to serve society rather than vice versa?

Crucially, coupling imagining with infrastructuring grounds it,
providing a context and a limit to claims. This prevents it from
becoming an arm-chair utopia in the worst sense, a naive project
detached from the constraints of reality and the current political
order. This materially-grounded imagining finds parallels in
practice-based research \citep{candy2019}, where the creation of artifacts
drives inquiry, and speculative design \citep{dunne2014} that
imagines future scenarios and steps through their social, cultural, and
political implications. It resembles utopia-as-method, characterized by
\citep{levitas2013} as a speculative sociology of the future. For Levitas
(2013), this method has three facets: the archeological, concerned with
unearthing and critiquing current models of society advanced by
politicians, states, and other actors; the architectural, imagining how
the world might be otherwise; and the ontological, thinking about the
social actors that occupy this alternative world. Experiments in
critical AI can examine these facets through productions of alternative
histories, mining the textual archeologies of language models to devise
alternate architectures and ontologies of sociality.

\subsection{\texorpdfstring{\textbf{Conclusion: From Viability to
Inspirability}}{Conclusion: From Viability to Inspirability}}\label{conclusion:-from-viability-to-inspirability}

When computation and socialism are brought together, it is typically to
focus on the well-known socialist calculation debate, a long-running
discussion in comparative economics about the role of information and
the viability of replacing or simulating the market \citep{hayek1940, cottrell1993, bunyk2023}. The
debate has been recently revived as AI and increased processing power
seem to unlock new capabilities \citep{pahl2023, boettke2023}. Given companies like Walmart and Amazon already allocate
resources at scale based on masses of consumer data, \citep{phillips2019} suggest socialism's historical failures were more a
result of authoritarian excess, limited computation, and imperfect data
sets than, as neoliberalism argued, any hard epistemic constraints.

Fundamentally, however, the debate quickly becomes a conceptual
cul-de-sac because of its framing: proponents must demonstrate
unequivocally a mechanism for replacing the market and centrally
coordinated resources on a global scale. Our intervention is not then
about proving the \emph{viability} of any particular economic model or
decision-making system, but about a contribution to
\emph{inspirability}. It is about opening up the space of possibilities
rather than implementing, testing, and ultimately dismissing them. The
point is not to model in fine-grained detail the inputs and outputs of
economic flows, but to cultivate alternative visions of the society and
economy. In short, the aim is generative rather than simulative,
speculation rather than stress-testing.

In this way, socialist infrastructuring and imagining comes closer to
science fiction or speculative fiction. Both genres, in their own ways,
seek to ``blueprint'' alternative societies --- using scenarios,
characters, and world-building to carefully step through the work
necessary to forge more emancipatory or egalitarian socioeconomic
systems. Bogdanov's \emph{Red Star} (1909) envisioned a
rationally-planned Martian economy---and inspired generations of Soviet
economists who aimed to emulate it. Spufford's \emph{Red
Plenty} creates fictional characters and fabricated incidents closely
based on real anecdotes and contemporary observations \citep{meek2010}. And \emph{Ministry of the Future} steps through how our
twinned ecologic and economic crises might be tackled in imaginative but
concrete ways. While these may be classed as utopic fiction, their
creators are highly aware of the tensions and antagonisms intrinsic to
human nature and relationships.

Such texts are powerful vehicles for generating imaginative and
innovative solutions. Collections like \emph{Economic Social Fictions}
\citep{davies2019} recognize this power, merging ``merging social science
with fiction, design with politics, scholarship with experimental
forms'' as a way to ``revitalize economic thinking.'' By jumping to different times, either the past or the
present, or packaging their insights as ``fiction,'' they allow us to
temporarily escape the ``real present'' and its stifled sense of
possibility.

In this sense, socialist infrastructuring may be conceived as an updated
version of ``machines for designing the future'' \citep{fisher2010}. By
training a language model on historical texts, our synthetic Allende and
Beer bots enable conversations that are speculative but also grounded in
a set of real-world writings, experiences, and confrontations. Whether
adapting fiction generators or ``misusing'' generative AI models, we
could envision any number of like-minded projects that aim to achieve
similar ends. Such projects leverage computational affordances to break
out of the labyrinth of capitalist realism that critics have diagnosed.

Socialist infrastructuring makes a contribution on two levels. First,
it does not claim to prove itself as a totalizing vision or universal
``solution,'' but instead aims to explore how alternative logics might
be taken up at particular scales, for particular peoples and particular
use-cases. Importantly, this is a ratcheting down of scope not
radicality. Indeed, this is how ``real-world'' systems are
conceptualized, developed, and produced. Second, it does not need to
provide launch-ready software which caters for every case and
contingency, but aims to demonstrate that other approaches to ordering
socio-economic life are possible. This in itself is a worthwhile
``outcome.'' As \citep{pahl2023} stress, such experimentation is first
and foremost about ``the aesthetics of societal visions, about the
critical reflection of latent tendencies, pre-analytical visions and
imaginaries.''

While socialist infrastructuring has been used towards imagining
alternative socioeconomic futures, this is only one possible topic of
many. We see strong potential, for instance, in using this approach to
explore alternative systems in the context of the climate crisis, given
how clearly inadequate our existing paradigms are. How might a
computational system ``solve'' for an economic order which aimed not to
maximize GDP but to minimize GHG (greenhouse gas) consumption? As
bestselling books and scholarly articles in degrowth, green growth, and
ecosocialism attest \citep{saito2023, saito2024}, there is strong and rising
interest in these possibilities. Indeed, very recent research
\citep{schlichter2024} takes early steps in this direction, adopting a
Cybersyn-inspired approach to climate change and arguing for a radical
pivot from a capitalist system to one that acknowledges planetary
boundaries.

As with any approach, there are certainly inherent limits to this
methodological and conceptual route. Firstly, there is some basic
technical literacy required to set up software, experiment with
technical systems, or prompt models --- though we suggest, like the
original Cybersyn project, that a surprising amount can be achieved with
relatively modest technical means. Secondly, this approach aims for
imagination rather than operationalization. This is not designed for
those seeking to transform socialist paradigms into concrete products or
shovel-ready projects, but rather best suited to those caught within
existing systems that are seeking something at once more modest and more
challenging: to simply imagine how our institutions, economies, and
everyday life could be arranged in any other way. If our economic and
ecological crises stem (partially) from our crisis of imagination \citep{ghosh2017},
we need new tools to break this deadlock and offer an expanded
possibility-space.

\end{multicols*}

\bibliographystyle{plainnat}
\bibliography{refs}

\end{document}